\documentclass[prc]{revtex4}
\usepackage{graphicx}
\begin{document}
\title{Refractive Distortions of Two-Particle Correlations from Classical Trajectory Calculations}
\author{Scott Pratt}
\affiliation{Department of Physics and Astronomy,
Michigan State University\\ East Lansing, Michigan 48824}
\date{\today}

\begin{abstract}
Calculations of two-particle correlations usually assume particles interact only pair-wise after their final collisions with third bodies. By considering classical trajectories, we show that interactions with the mean field can alter the spatial dimensions of the outgoing phase-space-density profiles by tens of percent, consistent with more complicated quantum complications.
\end{abstract}

\maketitle

\section{Introduction}
\label{sec:intro}

Two-particle correlations provide insight into the space-time development of relativistic heavy-ion collisions. Specifially, the six-dimensional correlation function $C({\bf P},{\bf Q})$, measured as a function of the total and relative momenta, provides information about the the shape of the outgoing phase space density distribution for particles with momentum ${\bf P/2}$. In addition to the size and lifetime of the source, the shape of the phase space packet is also affected by the mean-field traversed by the particles after their last collision. Recently, these distortions were calculated assuming particles passed through a time-independent optical potential \cite{Cramer:2004ih,Miller:2005ji}. If the pions left a region of lower mass, it was found that the lensing effect of the potential was to distort the extracted source dimensions by a few tens of percent. The effect was analagous to lensing effects from Coulomb mean fields, which were shown to explain different apparent source sizes for positive and negative pions at AGS energies \cite{Barz:1998ce,Barz:1996gr}.

The refractive corrections studied in Reference \cite{Cramer:2004ih,Miller:2005ji} allowed for a more physically interpretation of the correlation data. When ignoring the corrections it appears that the RHIC fireball grows to a transverse (to the beam) radius of $\approx 13$ fm, expanding at a speed $\approx 0.7c$, and rapidly disintegrates at a time of $\approx 10$ fm/$c$ \cite{Retiere:2003kf}. The puzzling aspect of this picture is that the fireball surface must expand at a speed of 0.7$c$ from the initial collision to grow from its initial size of 6 fm to 13 fm in 10 fm/$c$, not allowing any time for the matter to accelerate. Correlation analyses typically provide three dimensions, $R_{\rm long}$, the longitudinal size defined parallel to the beam, $R_{\rm out}$, the dimension of the phase space packet perpendicular to the beam and outward along the direction of the pair's momentum, and $R_{\rm side}$, the sideward dimension which is perpendicular to both the pair momentum and the beam axis. The apparent sideward dimension in \cite{Cramer:2004ih,Miller:2005ji} was shown to be increased by the refractive effects of the potential, which would suggest the true radius of the fireball was somewhat smaller than the 13 fm previously believed. Reducing the size by one or two fm would provide a more physically plausible picture of the reaction's evolution. 

The calculations of Ref. \cite{Cramer:2004ih,Miller:2005ji} involved solving for single-particle outgoing wave functions in the presence of complex optical potentials. The wave functions are then symmetrized and the interference term provides the leverage for relating the spatial characteristics of the source to the measured correlation function. Quantum calculations such as these have some drawbacks. First, they are somewhat difficult to interpret due to the inherent complexity of solving for wave functions. Secondly, implementation becomes complictated if one were to account for time-varying potentials, or potentials without the spherical and boost symmetries assumed in \cite{Cramer:2004ih,Miller:2005ji}. Finally, it is difficult to extend such calculations beyond the case of identical particles as one then needs to consider the quantum three-body problem. For these reasons, it is important to understand the validity which these distortions can be calculated classically. Given that the inferred diameters of the RHIC fireballs are often near 25 fm, one might expect that classical considerations could be valid except at very low $p_t$. 

In the next section we review the relation between the measured correlation function $C({\bf P},{\bf Q})$, the emission probability $s(p,x)$ and the asymptotic phase space density profile, $f({\bf p},{\bf r},t\rightarrow\infty)$. By utilizing the connection between the phase-space density and the measured correlation, it is shown how classical calculations of the single-particle phase space density, which can be readily adapted to include mean-field effects, can be used to generate correlations. To provide a physical explanation of how mean-field effects alter the final phase space density and therefore the correlation function, a simple analytic model of a static cylindrical source is presented in Sec. \ref{sec:cylinder}. Section \ref{sec:qclass} presents a direct comparison of a quantum calculation with one based on classical trajectories. Collective flow plays a critical role in correlation phenomenology. A more sophisticated model incorporating longitudinal and radial flow is analyzed in Sec. \ref{sec:trajectory}. This paper focuses on the distortion of the sideward dimension, which is mainly due to refraction, but the apparent outward dimension is also affected. Various issues concerning the outward dimension are briefly discussed in Sec. \ref{sec:outward}, and conclusions are present in Sec. \ref{sec:conclusions}.

\section{Theory Background}
\label{sec:theory}

For non-interacting identical particles, two-particle probabilities can be calculated in terms of one-body source functions \cite{Lisa:2005dd}. Assuming small relative momentum this can be expressed as
\begin{equation}
\label{eq:master}
\frac{dN}{d^3p_ad^3p_b}=\frac{dN}{d^3p_a}\cdot\frac{dN}{d^3p_b}
+\left|\int d^4x_a s(P/2,x_a) e^{i(p_a-p_b)\cdot x_a}\right|^2,
\end{equation}
where the source function $s(p,x)$ represents the probability of emitting a particle of momentum ${\bf p}$ from space-time point $x$. The correlation due to the interference is the ratio of the first two terms. The source functions can defined quantum-mechanically in terms of the quantum ${\cal T}$ matrices,
\begin{equation}
\label{eq:sdef}
s(p,x)=\sum_i \int d^4\delta x {\cal T}^*_i(x+\delta x/2)T_i(x-\delta x/2)
e^{ip\cdot \delta x}.
\end{equation}
The matrix elements ${\cal T}_i(x)$ represent the amplitude for a particle to have its last interaction with the source at $x$ while the remainder of the source evolves into state $i$. The last interaction includes interaction through the mean field as Eq.s (\ref{eq:master}) and (\ref{eq:sdef}) remain valid in the presence of a mean field, although it is clearly difficult to understand how the mean field will alter the ${\cal T}$ matrices and therefore the source functions. For small relative momentum, one can make the smoothness approximation, 
\begin{equation}
s(P/2,x)S(P/2,y)\approx s(E_{{\bf P}/2},{\bf P}/2,x)s(E_{{\bf P}/2},{\bf P}/2,y),
\end{equation}
which allows one to write the correlation in terms of the outgoing phase space density for particles with momentum ${\bf P}/2$.
\begin{eqnarray}
\label{eq:cf}
C({\bf P},{\bf Q})&=&1+\int d^3r {\cal S}_{{\bf P}}({\bf r})
\cos({\bf Q}'\cdot{\bf r}),\\
\nonumber
{\cal S}_{\bf P}({\bf r})&=&
\frac{\int d^3r'_a d^3r'_b
f({\bf P}'/2,{\bf r'_a},t')f({\bf P}'/2,{\bf r}'_b,t')
\delta({\bf r}'_a-{\bf r}'_b-{\bf r})}
{\int d^3r'_a d^3r'_b
f({\bf P}'/2,{\bf r'_a},t')f({\bf P}'/2,{\bf r}'_b,t')},
\end{eqnarray}
where the primes denote the positions measured in the pair frame, and ${\bf Q}'={\bf p}'_a-{\bf p}'_b$ is the relative momentum in that frame. Although ${\cal S}$ is often called a source function it is more accurately a measure of the outgoing phase space distribution for particles of a given momentum.

An alternative approach is to separate out interaction with the mean field,  define the source function so that it describes the points at which particles had their last interaction other than through the mean field, then use the outgoing wave function, which is a solution to the equations of motion using the mean field, to describe the evolution from $x$ to the asymptotic momentum state \cite{Cramer:2004ih,Miller:2005ji,Barz:1998ce}. This involves replacing the phase factor used to describe the evolution from $x$ to its asymptotic state ${\bf p}
$ in the derivation of Eq. (\ref{eq:master}),
\begin{equation}
e^{i(p_a-p_b)\cdot x}\rightarrow \phi^*(p_a,x)\phi(p_b,x).
\end{equation}
The altered expression is then
\begin{eqnarray}
\label{eq:cm}
\frac{dN}{d^3p_ad^3p_b}&=&
\int d^4x ~\tilde{s}(p_a,x)|\phi(p_a,x)|^2\int d^4x ~\tilde{s}(p_b,x) |\phi(p_b,x)|^2\\
\nonumber
&+&\left|\int d^4x ~\tilde{s}(P/2,x)
\phi^*(p_a,x)\phi(p_b,x)\right|^2,
\end{eqnarray}
where $\phi$ is the outgoing single-particle wave function describing evolution through the mean field. The source functions $\tilde{s}(p,x)$ now refer to the points where a particle had its last non-mean-field interaction.

The motivation for writing these equations is to stress the equivalence of Eq. (\ref{eq:master}) where the effects of the mean field are incorporated into the source function and Eq. (\ref{eq:cm}) where the effects are included by altering the outgoing evolution operator. Since both approaches are equivalent, Eq. (\ref{eq:cm}) can equally motivate the description in terms of the outgoing phase space density, Eq. (\ref{eq:cf}). In this paper we consider applications of Eq. (\ref{eq:cf}) where the phase space density is calculated by considering classical trajectories through the mean field. 

\section{Emission from a cylinder}
\label{sec:cylinder}

\begin{figure}
\centerline{\includegraphics[width=0.25\textwidth]{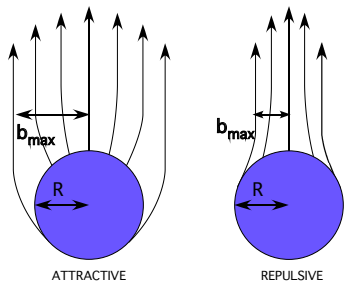}}
\caption{\label{fig:cartoon}
Trajectories that lead to the same asymptotic momentum are altered by an attractive mean field (left). The phase space distribution for particles of this given momentum are widened in coordinate space by a factor $b_{\rm max}/R$. Repulsive mean fields (right) reduce the size of the region of the outgoing particles.
}
\end{figure}

As a simple example we consider a pion being emitted from the surface of a cylinder of radius $R$, where the in-medium mass of the pion at the surface is $m_{\rm med}$ and the asymptotic vacuum mass is $m_{\rm vac}$. We assume the mass returns to its vacuum value exponentially,
\begin{equation}
m^2(r)=m_{\rm vac}^2+(m_{\rm med}^2-m_{\rm vac}^2)e^{-(r-R)/a}.
\end{equation}
From time-reversal arguments a thermal source will emit particles of given momentum with the same trajectories as those that describe absorption. The asymptotic trajectories of particles with momentum $p_x=p_t$, $p_y=0$, that intersect with the cylinder have a uniform distribution of impact parameters up to a maximum $b$ as illustrated in Fig.~\ref{fig:cartoon}. The effect of an attractive mean field is to stretch the width of the phase space cloud by a factor $b_{\rm max}/R$. The sideward dimension measured in two-particle correlations is stretched by this factor.

The maximum impact parameter for capture can be found by combining conservation of energy and angular momentum, $L=bp_t$,
\begin{eqnarray}
\label{eq:econs}
p_t^2&=&p_r^2+V_{\rm eff}(r)\\
\nonumber
V_{\rm eff}(r)&=&\frac{b^2p_t^2}{r^2}+m^2(r)-m_{\rm vac}^2.
\end{eqnarray}
The effective potential will have a maximum outside $R$ unless the momentum is small or if $a$ is large. The condition for maximum impact parameter $b_{\rm max}$ is that at the maximum of $V_{\rm eff}$ defined by
\begin{eqnarray}
\label{eq:cond1}
\frac{dV_{\rm eff}}{dr}&=&0\\
\nonumber
&=&-\frac{b_{\rm max}^2p_t^2}{2r^3}+[m_{\rm vac}^2-m^2(r)]/a,
\end{eqnarray}
the radial momentum is zero so that Eq. (\ref{eq:econs}) becomes
\begin{equation}
\label{eq:cond2}
p_t^2=\frac{b_{\rm max}^2p_t^2}{r^2}+m^2(r)-m_{\rm vac}^2.
\end{equation}
Equations (\ref{eq:cond1}) and ({\ref{eq:cond2}) can be combined to eliminate $m^2$ and provide a cubic equation for $r$ as a function of $b$. After finding $r$, one can use either equation to then find $p_t$ as a function of $b$. For higher $p_t$, the location of the minimum moves inside $R$ which allows one to ignore Eq. (\ref{eq:cond1}) and solve Eq. (\ref{eq:cond2}) with $r=R$. 
\begin{equation}
b_{\rm max}^2=R^2\frac{m_{\rm vac}^2-m_{\rm med}^2}{p_t^2}, \ \ \ 
p_t^2>(m_{\rm vac}^2-m_{\rm med}^2)[(R/2a)-1].
\end{equation}
The high $p_t$ behavior is independent of $a$, and in the case where $a>R/2$, the result is independent of $a$ for all $p_t$.

\begin{figure}
\centerline{\includegraphics[width=0.45\textwidth]{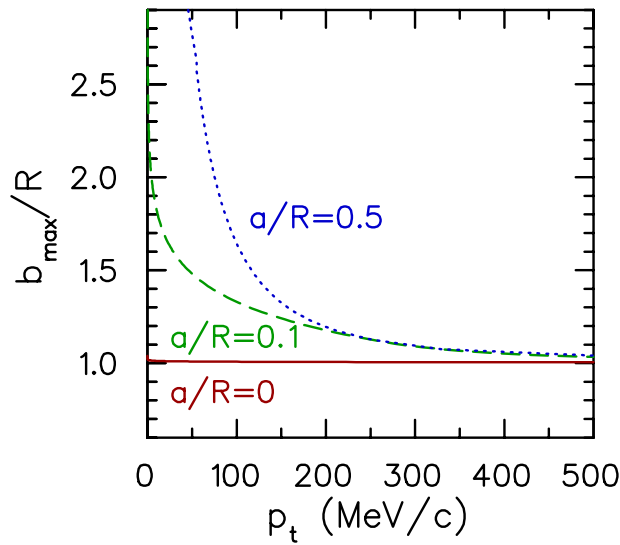}}
\caption{\label{fig:cylinder}
Assuming a static cylindrical source, the distortion to the sideward dimension due to the mean field is shown for attractive scalar fields. The field lowers the pion mass to 50 MeV/$c^2$ and falls off exponentially outside the emitting radius $R$ with a distance scale $a$. The distortion is stronger for larger ranges $a$ and at lower $p_t$.}
\end{figure}

Figure \ref{fig:cylinder} shows the ratio $b_{\rm max}/R$ as a function of $p_t$ for several values of $a$ given the case of a lighter in-medium pion mass, $m_{\rm med}=50$ MeV/$c^2$. For $a=0$, the mean field has no effect as the capture cross section does not extend beyond the cylinder. For the saturating value, $a=R/2$, the cross section is significantly enhanced. 
For $a=R/10$, the enhancement is identical to the saturating value at large $p_t$ and is somewhat reduced at low $p_t$.

The enhancement of the apparent sideward size is characteristic of an attractive mean field. This is consistent with Liouvilles theorem, which states that contraction of the phase space density in momentum space as it leaves the attractive mean field should be accompanied by a growth of the phase space cloud in coordinates space. Another example of an attractive mean field is the Coulomb field for the residual nucleus which attracts negative pions, but repels positive pions. This problem was addressed by Barz, who performed quantum calculations using Coulomb wave functions to describe the evolution through the mean field \cite{Barz:1996gr,Barz:1998ce}. The effect was shown to be significant at AGS energies, where it explained the large observed differences between apparent sizes for positive and negative pions.

\section{Direct Comparison of Quantum and Classical Calculations}
\label{sec:qclass}

For potentials where all the characteristic length scales far exceed $\hbar/p_t$, one expects that classical and quantum calculations should converge. The radius of the RHIC fireball is close to 10 fm, which readily satisfy the condition until $p_t$ is below $\sim 50$ MeV/c. However, there are other length scales, such as the diffuseness parameter that describes how quickly the density falls off a the surface. We consider an optical potential with a Fermi-Dirac form,
\begin{equation}
\label{eq:optical}
U(r)=(U_R+iU_I)\frac{1}{e^{(r-R)/a}+1},
\end{equation}
where $a$ is the diffuseness parameter. If $a=0$, the potential is a step function. Quantum wave packets can reflect off step-function potential barriers even when transmission is energetically allowed. This is also true for imaginary potential barriers. As an extreme example, a classical particle that heads into a barrier, $U(x)=iU_I\Theta(x)$, with $U_I\rightarrow\infty$ will be completely absorbed by the barrier, whereas a quantum wave will be completely reflected. Thus, we expect the quantum-classical comparison to depend on the sharpness of the potential barrier.

In order to quantitatively assess the validity of classical trajectory calculations as an alternative to quantum calculations, we consider a static two-dimensional complex optical potential as described in Eq. (\ref{eq:optical}) and solve for the distortion to the apparent sideward radii for particles which originate isotropically from points a radius $r_0$ from the center of the cylinder. For identical particles, radii can be determined from correlations using the expression \cite{Lisa:2005dd},
\begin{eqnarray}
\label{eq:hbtmoments}
\langle x_ix_j\rangle&=&
\left.\frac{1}{2}\frac{d^2C({\bf P},{\bf Q})}{dQ_idQ_j}\right|_{{\bf Q}=0},
\end{eqnarray}
where the radii represent the dimensions of the asymptotic phase space density, not the dimensions of the emission points. Using Eq. (\ref{eq:cm}) which gives the correlation function in terms of the distributions points for last collisions convoluted with outgoing wave functions, one can then apply Eq. (\ref{eq:hbtmoments}) to write an expression for the variance of the sideward size,
\begin{equation}
R_{\rm side}^2\equiv\langle y^2\rangle
=\frac{\int d^4x \tilde{s}(p,x) \frac{d}{dp_y}\phi^*({\bf p},x)
\frac{d}{dp_y}\phi({\bf p},x)}
{\int d^4x \tilde{s}(p,x)\phi^*({\bf p},x)\phi({\bf p},x)},
\end{equation}
where the asymptotic momentum ${\bf p}$ moves along the $x$ axis.

The wave functions $\phi({\bf p},x)$ were calculated numerically by integrating the Klein-Gordon equation for outgoing and incoming cylindrical partial waves from $r$ outside the range of the potential to $r=0$. Outside the range of the potential the partial waves have the form,
\begin{equation}
\phi_{\rm ell,\pm}(r)\rightarrow \frac{1}{2}
(J_\ell(p_tr/\hbar)\pm iY_\ell(p_tr/\hbar)), 
\end{equation}
After integrating the Klein-Gordon equation to find the incoming and outgoing solutions at all $r$, one makes the combination
\begin{equation}
\phi_{\ell}(r)=\phi_{\ell,+}(r)-\beta\phi_{\ell,-}(r),
\end{equation}
where $\beta$ is chosen to satisfy the boundary condition, $\phi_\ell(r=0)=0$.  The outgoing plane wave is then
\begin{equation}
\phi({\bf p},{\bf r})=\phi_0(r)+\sum_{\ell=1,\ell_{\rm max}}2i^\ell \cos(\ell\phi)
\phi_\ell(r).
\end{equation}
Depending on the size of $p_tr/\hbar$, the sum over angular momenta might extend to $\ell_{\rm max}$ above 50 before results would converge.

\begin{figure}
\centerline{\includegraphics[width=0.5\textwidth]{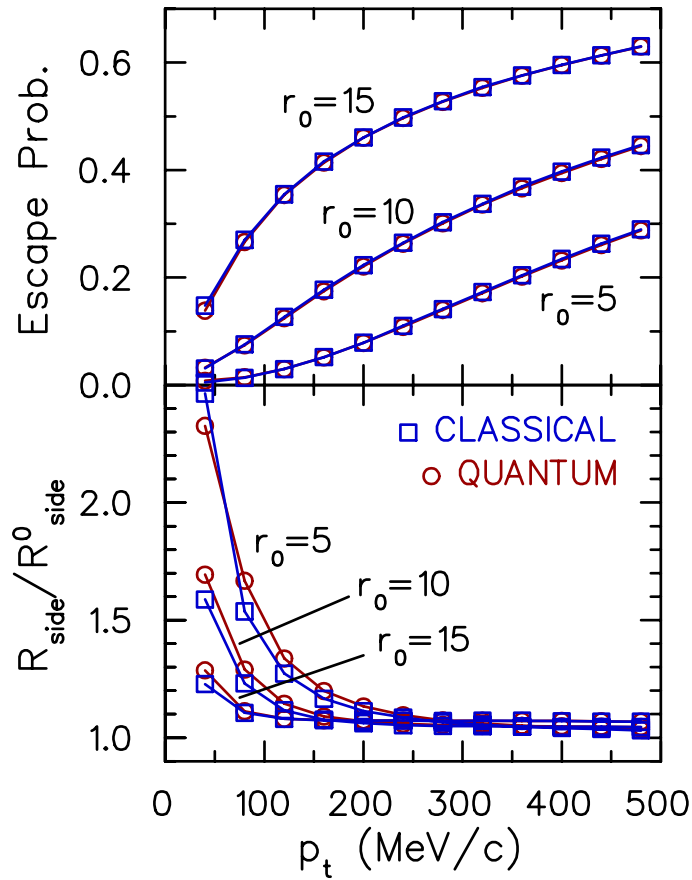}}
\caption{\label{fig:qclass}
The ratio of the apparent sideward dimension to the dimension without mean field is shown as a function of $p_t$ for the optical potential described in the text. Distortions are calculated for both quantum calculations (circles) and classical trajectory calculations (squares). The source function describing the final collision points was confined to an intial radius $r_0$=5,10 or 15 fm. The upper panel shows the probability that such particles escape without being absorbed due to the imaginary part of the optical potential. The two approaches agree within a few percent.}
\end{figure}

Calculations were performed for $\tilde{s}$ being independent of the direction of ${\bf p}$ with emission points being confined to a radius of $r_0$. The potential parameters were $U_R=m_{\rm med}^2-m_{\rm vac}^2$ with the in-medium mass $m_{\rm med}=50$ MeV$/c^2$ and $U_I=m_\pi\cdot\Gamma_0$ with $\Gamma_0=100$ MeV. This corresponds to a classical decay rate of $\Gamma=\Gamma_0 m_\pi/E$. The fact that the decay rate falls $\sim 1/E$ is characteristic of a scalar form for the optical potential. The lower panel of Fig. \ref{fig:qclass} shows the distortion of $R_{\rm side}$, defined as the ratio of $R_{\rm side}$ with the potential to $R_{\rm side}$ with $U=0$, as a function of $p_t$. The distance scales for the potential were $R=10$ fm and $a=3$ fm. Results are shown for three different values of $r_0$, 5 fm, 10 fm and 15 fm. The distortions rise at low $p_t$ with stronger distortions for small $r_0$. Also displayed in Fig. \ref{fig:qclass} are escape probabilities. Quantum mechanically, the probability that a particle escapes the fireball is
\begin{equation}
P_{\rm escape}=\frac{\int d^4x \tilde{s}(p,x) |\phi({\bf p},{\bf r})|^2}
{\int d^4x\tilde{s}(p,x)}.
\end{equation}
Since low $p_t$ particles spend more time in the fireball, and since the decay rate falls as $1/E$, escape rates are small at low $p_t$.

The corresponding classical calculations were performed by solving for the trajectories of particles through the mean field. The initial points were chosen on a circle of radius $r_0$ with random directions and the energy chosen so that their final momentum would have magnitude $p_t$. After the trajectory was calculated, phase space points were rotated to make $p_y=0$ and $\langle y^2\rangle$ was calculated. Trajectories were calculated consistent with the relativistic equations of motion,
\begin{equation}
\frac{d{\bf p}}{dt}=-(1/2E)\nabla ~\Re U(r),
\end{equation}
while particles were decayed randomly throughout the trajectory with a rate $\Gamma=(1/E)~\Im U(r)$. Classical calculations are shown alongside the corresponding quantum calculations in Fig. \ref{fig:qclass}.

The apparent sideward source sizes from the calculations used for Fig. \ref{fig:qclass} agree well with one another, differing by only a percent or two for $p_t>100$ MeV/$c$. Even at $p_t=40$ MeV/$c$, the calculations agree to better than 10\%. As explained above, the agreement should be expected to worsen for smaller $a$. Repeating the calculations for $a=0.5$ fm, discrepancies increased to the level of a few percent for $p_t>100$ MeV/$c$, and notably higher for $p_t\sim 50$ MeV/c. However, it is hard to physically motivate such a sudden change in densities.

\section{Cylindrical Geometry with Collective Expansion}
\label{sec:trajectory}

The simple cylindrical picture of the previous section ignored collective radial and longitudinal expansion. As collective expansion strongly influences both correlation measurements and spectra, it is important to understand how mean-field distortions are affected by both radial and longitudinal expansion. Here, we consider a cylinder where particles are emitted from a surface of radius $R$ at a time $\tau=\sqrt{t^2-z^2}$. The matter has a boost-invariant collective velocity along the beam axis, $v_z=z/t$, and a transverse collective rapidity $y_t$ at radius $R$. The break-up surface is allowed to move at a different transverse rapidity $y_b$. This surface should be moving outward, $y_b>0$, at early times when the system is expanding to reach its maximum radii which are probably near 13 fm. The breakup surface should then stop and collapse with $y_b$ becoming negative, and might even become time-like. Thus, the parameters describing the expansion, $R$, $y_t$, $y_b$ and the temperature $T$, should be chosen based on which phase of the emission one wishes to study.

As in the previous section, we consider a scalar mean field where the pion mass has the form,
\begin{equation}
m^2(r)=m_{\rm vac}^2+(m_{\rm med}^2-m_{\rm vac}^2)e^{-(r-R+v_bt)}.
\end{equation}

The initial momenta of the particles are generated randomly to be consistent with a thermal distribution with an outward boost of rapidity $y_t$. In addition to the Boltzmann weight, an extra weight proportional to the velocity relative to the boundary is included so that the initial generation of phase space points is consistent with the flux of a thermal distribution through the boundary. If at any time a trajectory re-enters the interior of the boundary, the trajectory is discarded. All trajectories are generated from the same initial point, $x=R, y=z=0, t=\tau$. After the trajectory has been calculated, rotational and boost invariance are used to boost and rotate the trajectory so that the asymptotic momenta satisfy $p_y=p_z=0$. The trajectories are then binned as a function of the asymptotic transverse momentum, $p_x=p_t$, and the average and variance of the coordinates $x,y,z$ at an asymptotic time $t$ are calculated for each $p_t$ bin. Since all particles within a given bin have the same velocity, the variances are independent of the asymptotic time. 

The variance of the $z$ dimension can be identified as $R_{\rm long}$ while the variance of $y$ dimension can be identified as $R_{\rm side}$. Since emission occurs for a variety of radii and times, the experimentally inferred dimensions should be generating by averaging over various values for $T,R,\tau,y_t$ and $y_b$. For the purposes of gaining insight into the distortions, we study the distortions for specific choices of the parameters. The relative weight of emission during the expansion stage, where $y_b>0$ and the final burst where $y_b<0$, would depend on details of the equation of state. The variance of the $x$ coordinate is also sensitive to the relative times of the emission stages, and we avoid discussing the distortion to the outward dimension $R_{\rm out}$.

\begin{figure}
\centerline{\includegraphics[width=0.6\textwidth,angle=90]{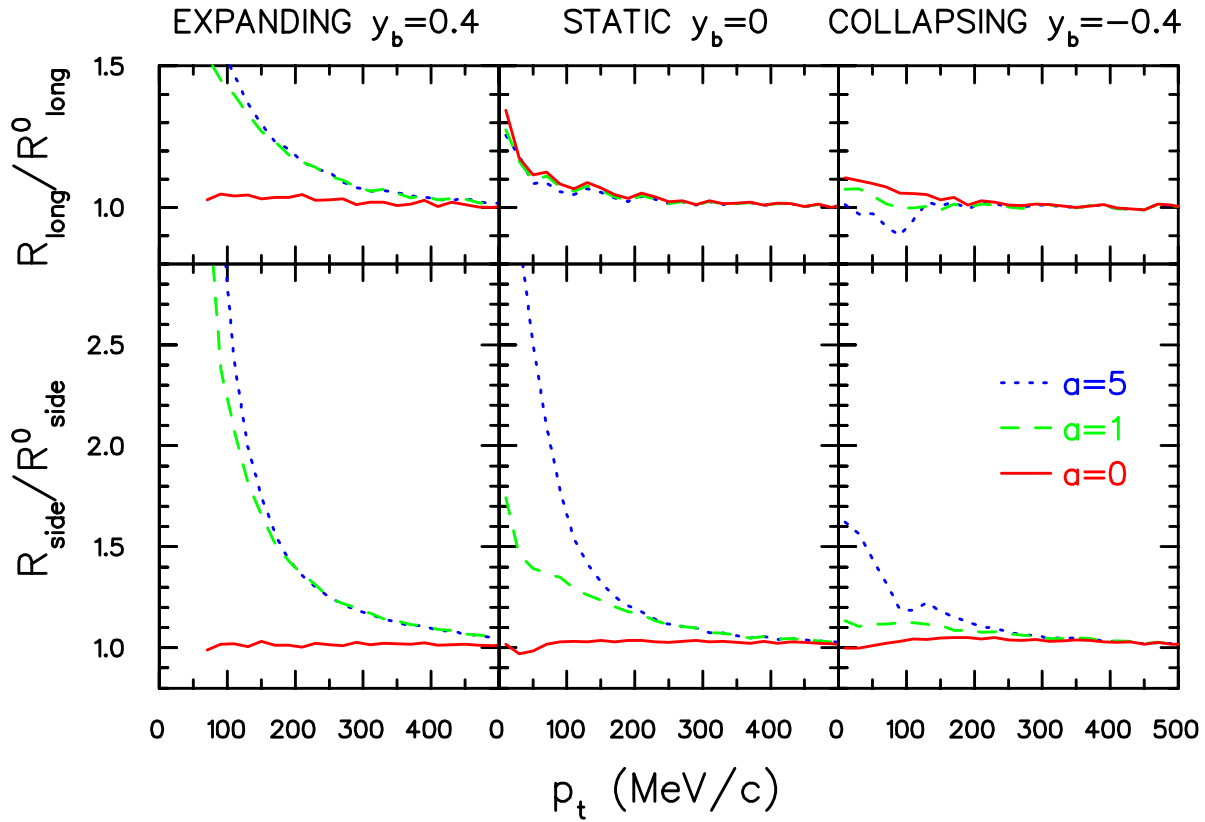}}
\caption{\label{fig:trajectory}
The distortion of the sideward and longitudinal dimensions due to an attractive scalar mean field are shown as a function of transverse momentum. Calculations involved solving classical trajectories through a mean field where the pion mass at the breakup surface was 50 MeV/$c^2$ and returned to the vacuum value exponentially relative to the moving surface. Results are shown for three values of the exponential scale $a$. Three sets of parameters are chosen to represent the system while it is: expanding (left panels, $v_b=0.4, R=9$ fm, $\tau=9$ fm/$c$), has reached its maximum (center panels, $v_b=0, R=12$ fm, $\tau=12$ fm/$c$), and is collapsing (right panels, $v_b=-0.4, R=9$ fm, $\tau=15$ fm/$c$). For all calculations momenta were initialized according to a temperature of 120 MeV and a collective transverse rapidity of 0.8. The distortion is small if the range over which the mean field is non-zero is small $(a=0)$, and becomes very significant when the range approaches 5 fm. Distortions are more pronounced when the breakup surface is moving outward as the particles then have more opportunity to interact with the mean field.}
\end{figure}

Trajectories were calculated with and without the mean field. Distortions of the radii are presented in Fig. \ref{fig:trajectory} by showing the ratios $R_{\rm side}$ and $R_{\rm long}$ to the dimensions calculated without the mean field. For each case the ratios are shown for one temperature, $T=120$ MeV, and for three values of $a$: 0, 1 fm and 5 fm. For each choice of parameters, one million trajectories were calculated, which was sufficient to suppress statistical fluctuations except at $p_t$ below 100 MeV/$c$ where the $p_t$ bins are not well populated.

The left panels of Fig. \ref{fig:trajectory} show results for an expanding system where both the boundary and the medium have positive outward velocities. The surface was assumed to be at a radius of 9 fm at a time $\tau=9$ fm/$c$ while moving at a transverse rapidity of $y_b=0.4$. The thermal distribution was characterized by a collective transverse rapidity of $y_t=0.8$. For these parameters, the distortions for $R_{\rm side}$ were even stronger than they were for the static cylinder from the previous section. This owes itself to the fact that with $y_b>0$, the forces from the potential follow the particles and push them for a longer time. For this case there is also significant distortion of the longitudinal dimension. The center panels illustrate the distortion for particles created  when the breakup surface has reached its maximum, $y_b=0$, with the matter rapidity again chosen to be $y_t=0.8$. Here the radius and time were both assumed to be 12 fm and 12 fm/$c$. The distortion of the longitudinal dimension is small while the distortion of the sideward dimension very much mimics the results for the static cylinder of Sec. \ref{sec:cylinder}. Finally, the right panels show representative results for when the breakup surface is collapsing. Parameters for this case were $R=9$ fm, $\tau=15$ fm/$c$, $y_b=-0.4$ and $y_t=0.8$. Here, the distortions are smaller than for the previous two cases since the particles spend less time under the influence of the mean field.

The results of Fig. \ref{fig:trajectory} can not be directly compared to those of Ref. \cite{Cramer:2004ih,Miller:2005ji} since those calculations used a continuous distribution of initial radii modified by an escape probability determined by the imaginary part of an optical potential. In their calculation, the in-medium pion mass was also 50 MeV/$c^2$, but since particles tended to be emitted away from the center, the mass at the average emission point would be larger. Thus, it is not surprising that the distortions of Fig. \ref{fig:trajectory} are somewhat larger, by 50-100\%, than those of reference \cite{Cramer:2004ih,Miller:2005ji}. The shape of the $p_t$ dependence and the relatively larger distortion of $R_{\rm side}$ vs that of $R_{\rm long}$ were similar for both calculations.

\section{The Outward Dimension}
\label{sec:outward}

The distortion of the apparent outward size was neglected in this study. The reason for restraining from reporting these distortions comes from the nature of the illustrative models applied in the previous sections. In these models emission from a specific radius was considered to provide a better understanding of the distortion. Since the outward dimension is affected by the duration of the emission, the results for the outward dimensions from these calculations, where the radius and emission time are fixed, are unphysical. For instance, if one has a large imaginary potential and effectively emits from only the surface, one needs to consistently incorporate the sequential emission from the excited regions blocked by the absorptive potential. Thus, any claims about the effect of the real and imaginary potentials on $R_{\rm out}$ can easily be misleading. Nonetheless, we wish to delineate three ways in which the outward size will be affected by such potentials.

(1) A strong imaginary potential confines emission to the surface. Although this might lead to a reduction in the $R_{\rm out}$ parameter in a toy model, the opposite trend is likely to ensue for consistent dynamical calculations. Pions with $p_t>100$ MeV/$c$ move faster than the expansion velocity of the fireball's surface. By delaying the emission of pions from the interior of the fireball, the pions emitted from the surface get a {\it head start} which can lead to the phase space cloud being more extended in the outward dimension.

(2) The particle's velocities are reduced. The group velocity is affected by the medium in both classical and quantum calculations. For eikonal approaches such as those discussed in Ref.s \cite{Kapusta:2005pt,Wong:2004gm,Chu:1994de}, the interference between two paths from space-time points $x_1$ and $x_2$ to asymptotic momentum states $p_1$ and $p_2$ is governed by the phase,
\begin{eqnarray}
\label{eq:eikonal}
&&-i(E_1-E_2)(t_1-t_2) +i(p_1-p_2)(x_1-x_2) +i\delta(p_1,x_1)-i\delta(p_2,x_1)
+\delta(p_2,x_2)-\delta(p_2,x_1)\\
\nonumber
&&\hspace*{40pt}\approx i(p_1-p_2)\left\{[(x_1-v_pt_1 +(d\delta_1/dp)]
-[(x_2-v_pt_2 +(d\delta_2/dp)]\right\},
\end{eqnarray}
where the relative momentum is assumed to be small and the group velocity is $dE/dp$. Here, $\delta(p,x)$ is the eikonal phase found by integrating over the straight-line trajectory. The derivative of the phase shifts can be equated with time and spatial delays,
\begin{eqnarray}
\delta(p,x)&=&\int_x^\infty dx\left[p(x)-p\right],\\
\frac{d\delta}{dE}&=&\frac{1}{2}\int dx
\left[\frac{1}{v(x)}-\frac{1}{v}\right],
\end{eqnarray}
where $p(x)$ is the momentum of the particle when it passes point $x$ on its way to its asymptotic momentum $p$.
Since $dx/v$ is the time spent traversing the interval $dx$ for a classical trajectory, $d\delta/dE$ is the time delay induced by the potential and $v_pd\delta/dE=d\delta/dp$ is the spatial offset. Thus, the phase in Eq. (\ref{eq:eikonal}) can be expressed as:
\begin{equation}
i(p_1-p_2)[x_1(t\rightarrow\infty)-x_2(t\rightarrow\infty)],
\end{equation}
where $x_1-x_2$ is the asymptotic relative separation of two particles with the same momentum $(p_1+p_2)/2$ as calculated with classical trajectories. This demonstrates an equivalence (exact in the limit of small relative momentum) agreement the eikonal and classical trajectory approaches in a one-dimensional system. The classical perspective provides a physically transparent understanding of the distortions to the outward size from a mean field. For instance, the apparent outward size can be reduced if those particles emitted early, or further ahead, would be more strongly retarded by the mean field than those emitted later. 

(3) Curvature of the trajectories can lead to emission from regions further backward from the direction define by the momentum. For the case of an attractive potential illustrated in Fig. \ref{fig:cartoon}, some of the trajectories are allowed to originate from the edges, and in some cases from the back, of the fireball. For the instantneous-emission pictures considered in the previous sections, these distortions tended to extend $R_{\rm out}$ and $R_{\rm side}$ by similar factors. Classical Boltzmann equations can consistently incorporate all three of these effects.

\section{Conclusions}
\label{sec:conclusions}

The principal conclusion from this study is that the effects of a particle traversing a mean field between its last collision and when it reaches its asymptotic straight-line trajectory can be modeled by calculating classical trajectories, then using the final phase space distributions to generate correlation functions. This approach should be valid except at very low $p_t$, unless there is a very sharp discontinuity in the mean field. Classical calculations are more tenable than quantum distorted-wave approximations. They are simpler to perform and can be applied even when the fields are time-dependendent or when there are no symmetries to reduce the dimensionality of the wave-function solutions. Another advantage of the classical model is that they can be applied to the case of correlations generated by strong and Coulomb interactions by replacing the cosine term in Eq. (\ref{eq:cf}), which gives the correlation in terms of the asymptotic phase space density, with the squared relative wave function \cite{Lisa:2005dd}.

The magnitude of the distortions in Fig. \ref{fig:trajectory} are significant, of the order of tens of percent when $p_t\sim 100$ MeV/$c$. However, it should be noted that the conditions chosen for the trajectory calculations in Sec. \ref{sec:trajectory} were somewhat extreme. Not only was the in-medium mass chosen to be have changed by a large amount, the large change was applied at the breakup radius from which the particles suffered their last collision. Although this is a strong assumption, it is reasonable to assume that the real part of the potential extends beyond the radii at which final collisions occur. This is because the real part of the optical potential should be much larger than the imaginary part of the potential at low temperature. In RPA calculations, the potentials can be calculated in terms of the phase shifts calculated using the relative momenta typical of the medium's temperature. At temperatures below 100 MeV, the invariant-masses of the collisions are well below that required for the $\rho$ and $K^*$ resonances, and are beginning to fall below that required for the $\Delta$ resonance. The relative strength of the real and imaginary part of the potential is $\tan\delta$, and since the phase shifts fall well below 90 degrees at low temperature, the real parts dominate. It should be stressed that any attractive potential will lead to distortions that broaden the measured sizes. Whereas a falling pion mass is rather controversial, in the low density limit one can calculate the mean field with some confidence using measured pion-pion and pion nucleon phase shifts. At very low temperature, only $s$-wave phase shifts contribute and the interactions are both repulsive and attractive depending on the channel. For temperatures in the range of 100 MeV, $p$-wave interactions play an important role. These interactions tend to be strongly attractive due to level repulsion caused by mixing of the pion states with $\Delta$-hole, $\rho$-hole or $K^*$ states \cite{Koch:1992zi}. Such mixing was even used to motivate pion condensation at high baryon density \cite{Migdal:1978az}. Though such extreme modifications of the dispersion relation are unlikely given the small baryon density at breakup, the overall density of hadrons at breakup is probably  near normal nuclear density $\sim 0.15$ hadrons/fm$^3$, and the pion dispersion relation might be modified by several tens of MeV.

The calculations presented here were only meant to be illustrative. The potential magnitude of mean-field effects and the ability of classical pictures to model the effects underscore the importance of incorporating mean field effects into hadronic Boltzmann descriptions of the breakup stage. Such calculations have been undertaken in the past, especially for lower-energy Fermi-velocity collisions, where the mean field is paramount \cite{gong,Bauer:1993wq}. In this energy range, phase space points used to generate correlations are typically taken from when a particle leaves the region of mean field, rather than when they have their last collision. Given the likelihood that the dispersion relations for pions are non-trivial, it is important that calculations incorporate momentum-dependent interactions. Significant theoretical effort has already been applied at developing a theoretical framework for modeling low-density transport that consistently handles both collisions and the modification of the particle's group velocities \cite{Danielewicz:1995ay,Bass:1998ca,Morawetz:1998ev,Leupold:2000ma}. Although this development has largely been aimed at intermediate-energy physics, it should be quite applicable for the breakup phase at high-energy. Such an effort is crucially important for reconstructing the space-time evolution of the fireball, and for understanding the pressure in the hadronic phase.

\begin{acknowledgments}
Thoughtful and inspiring comments from G. Miller, J. Cramer and P. Danielewicz are gratefully acknowledged. Support was provided by the U.S. Department of Energy, Grant No. DE-FG02-03ER41259.
\end{acknowledgments}

\end{document}